\documentclass[preprint]{aastex}

\def\NHthree{NH$_3$}          % H3+
\def\NtwoHp{N$_2$H$^+$}        % N2H+ 
\def\thCO{$^{13}$CO}        % N2H+ 
\def\CeiO{C$^{18}$O}        % N2H+ 
\def\CsvO{C$^{17}$O}        % N2H+ 
\def\Ntwo{N$_2$}               % N2
\def\Htwo{H$_2$}               % N2
\def\av{$A_V$}               % N2

\begin{document}
\setcounter{figure}{0}

\title{ N$_2$H$^{+}$ and C$^{18}$O Depletion in a Cold Dark Cloud}

\author{Edwin A. Bergin}
\affil{Harvard-Smithsonian Center for Astrophysics, 60 Garden Street,
Cambridge, MA 02138} \email{ebergin@cfa.harvard.edu}

\author{Jo\~ao Alves}
\affil{European Southern Observatory, Karl-Schwarzschild-Strasse 2, 85748
Garching, Germany}
\email{jalves@eso.org}        

\author{Tracy L. Huard }
\affil{Harvard-Smithsonian Center for Astrophysics, 60 Garden Street,
Cambridge, MA 02138} \email{thuard@cfa.harvard.edu}

\author{Charles J. Lada}
\affil{Harvard-Smithsonian Center for Astrophysics, 60 Garden Street,
Cambridge, MA 02138} \email{clada@cfa.harvard.edu}

\slugcomment{accepted by the The Astrophysical Journal (Letters)}
\begin{abstract}
We present sensitive, high angular resolution molecular-line observations
of C$^{18}$O and \NtwoHp\ toward the dark globule B68. We 
directly compare these data with the near-infrared extinction measurements of 
\citet{alves_b68}
to derive the first evidence for the depletion of \NtwoHp , and by
inference \Ntwo , in  a pre-stellar dark cloud.  
%This evidence
%is found by comparison of molecular emission map to the distribution of 
%total column density derived from near-infrared extinction techniques
%by \citet{alves_b68}.   
We also find widespread C$^{18}$O depletion 
throughout the centrally condensed core of the B68 cloud. Specifically,
we find the \NtwoHp\ 
emission to peak in a shell partially surrounding the peak of dust extinction.
Moreover, the \NtwoHp\ peaks inside
the much larger \CeiO\ depletion hole and has a smaller depletion zone, confirming
theoretical predictions. 
These data are analyzed through a direct coupling of time dependent
chemical models to a radiation transfer code.  This analysis highlights
the importance of photodissociation at cloud edges and suggests that
the CO abundance declines by two orders of magnitude from edge to center.
In contrast \NtwoHp\ declines in abundance, at minimum, by at least a factor of
two.  Indeed it is entirely possible that both \NtwoHp\ and \Ntwo\ are 
completely absent from the central regions of the B68 core.
The depletion of \NtwoHp , and its parent molecule \Ntwo , opens the possibility 
that the centers of dense cores, prior to the formation of a star, 
may evade detection by conventional
methods of probing cores using molecular emission.    
Under these conditions H$_2$D$^+$ may be the sole viable molecular probe 
of the innermost regions of star forming cores.
\end{abstract}


\section{Introduction}

Over the past decade there has been converging observational and theoretical
evidence of differential molecular gas phase depletions
that tracks the dynamical evolution of  star forming cores.   
This chemical sequence 
begins with the depletion of sulfur-bearing molecules,
which have strong bonds to grain surfaces and
are predicted to show the largest depletion ``holes'' in their emission 
\citep[][hereafter BL97 and C97]{bl97, charnley_dep}; indeed 
shell-like structures are now commonly observed for such molecules
in low
mass cores \citep[e.g.,][hereafter Tf02]{kuiper_l1498, ohashi_ccs, tafalla_dep}.   
Later in the process even volatile molecules, such as CO, deplete; 
as clearly demonstrated by
several observational efforts \citep{alves_l977, kramer_ic5146,
caselli_l1544, jessop_l1689}. 

Besides \Htwo , the molecule predicted to be least affected by the 
condensation process is \Ntwo , which is more volatile than CO and is
easily removed from grain surfaces by whatever desorption process 
acts in molecular cores (BL97; C97).  The homonuclear molecule
\Ntwo\ is not directly observable, 
but its chemical daughter products, \NHthree\ and \NtwoHp , 
are observed to trace dense cloud cores where other species
such as CO and CS are depleted \citep[][Tf02]{bergin_ic5146},
consistent with theoretical expectations \citep[BL97, C97,][]{aikawa_dep, li_dep}.
%Thus theory has predicted a sequence of differential depletions from
%CS to CO and finally \Ntwo\ \citep[e.g.,][]{bl97, aikawa_dep, li_dep}, 
%which is borne out by observations.

In this letter we present evidence for the depletion of both \NtwoHp\ 
and \CeiO\ in the cold pre-stellar B68 dark cloud. 
%This is the first detection
%of \NtwoHp\ depletion in a dark molecular cloud core and was made possible by the 
%direct comparison of our sensitive molecular-line observations with the recent
%near-infrared extinction maps of the cloud made by 
%\citet[][hereafter ALL01]{alves_b68}.
The detection of \NtwoHp\ depletion opens the possibility that the cores of
molecular clouds may evade detection by conventional techniques which rely
upon molecules as probes.  We discuss the implications of this result
for future chemical and star formation studies.

\section{Telescope and Spectrometers}

The J=1-0 transitions of \CeiO\ (109.78218 GHz), \CsvO\ (112.359277 GHz),
and \NtwoHp\ (93.17378 GHz)
were observed during April 2000 and 2001 using the
IRAM 30m telescope.  \CeiO\ was mapped with Nyquist sampling,
whereas pointed observations of \CsvO\ were 
conducted using the \CeiO\ map as a guide.
For \NtwoHp\ the inner 50$''$ of B68 was mapped with Nyquist sampling, 
but the remainder of the map
was obtained with honeycomb (24$''$) spacing.  The half-power beam width at 110 GHz and
93 GHz is
22$''$ and 26$''$, respectively.    
The system temperatures were typically $\sim$160--190 K.
In April 2001 we also obtained observations of 
the J=3-2 transition of \NtwoHp\ (279.512 GHz; $\theta_{MB} \sim 10''$).
Each line was observed in frequency
switching mode with autocorrellators as the backends providing a velocity
resolution of 0.053 km s$^{-1}$ (\CeiO ), 0.02 km s$^{-1}$ (\NtwoHp\ 3-2),
and $\sim$0.1 km s$^{-1}$ (\NtwoHp\ 1-0, \CsvO ). 
Pointing was checked frequently on nearby point sources 
(in April 2001 Mars was located only 9$\arcdeg$ from B68) 
yielding an uncertainty of $\sim 3''$.  The data were 
calibrated using the standard chopper wheel method and are presented here
on the T$_{mb}$ scale using standard calibrations from IRAM documentation.    

%The J=1-0 transitions of \CeiO\ , \CsvO , and \NtwoHp\ 
%were observed during April 2000 and 2001 using the
%IRAM 30m telescope.  Relevant observational parameters are provided in Table~1.
%\CeiO\ was mapped with Nyquist sampling
%and pointed observations of \CsvO\ were 
%conducted guided by the \CeiO\ map.
%For \NtwoHp\ the inner 50$''$ of B68 was mapped with Nyquist sampling, 
%but the remainder of the map
%was obtained with honeycomb (24$''$) spacing.  
%In April 2001 we also obtained observations of 
%the J=3-2 transition of \NtwoHp .
%Each line was observed in frequency
%switching mode. Frequent pointing checks were performed 
%(in April 2001 Mars was located only 9$\arcdeg$ from B68) 
%yielding an uncertainty of $\sim 3''$.  The data were 
%calibrated using the standard chopper wheel method and are presented here
%on the T$_{mb}$ scale using standard calibrations from IRAM documentation.    

\section{CO and N$_2$ Depletion}

\subsection{Observations}

In Figure 1a and 1b we compare the integrated intensity distribution of \CeiO\ J=1--0 
and \NtwoHp\ J=1-0 (all hyperfine components) with
the map of visual extinction (\av ) obtained by \citet[][hereafter ALL01]{alves_b68}.  
The  \CeiO\ emission peaks in a partial shell-like structure with a
radius $\sim 50''$, approximately centered on the \av\ peak. 
Likewise, the \NtwoHp\ emission maxima form an arc centered on
the \av\ peak, but lie interior to the \CeiO\ shell.
%Both molecules
%have an extension towards the southeast, with \CeiO\ showing better agreement
%with the \av\ map in that direction.  
%Both \CeiO\ and \NtwoHp\ show 
%ordered line center velocity shifts 
%(of slightly less than a line width
%across the core) 
%and line width variations throughout
%the map.  Discussing these variations is beyond the scope of this letter and
%will be treated in detail in a separate publication \citep{lada_b68}.

In Figure 2a  we present a direct position-by-position
comparison to the visual
extinction data using the techniques first described in \citet{lada_ic5146}. 
For \CeiO\ there is little scatter with 
the main feature being a dramatic break in the linear rise in 
CO emission with visual extinction at \av\ $\sim 8$ mag. This break is followed by a 
slow but systematic {\em decrease} in CO emission at greater extinctions.
As discussed by \citet{lada_ic5146}, in the absence of
non-LTE excitation (as expected for \CeiO ), 
a linear correlation between integrated intensity and \av\ corresponds to
an constant abundance throughout the cloud.  A break in the correlation toward
greater extinction could be the result of either a decrease in molecular abundance
or high opacity saturating the molecular emission.   
%Using the
%ratio of \CeiO /\CsvO\ integrated intensities at the map center we derive
%a \CeiO\ opacity of 1.6 (assuming $^{18}$O/$^{17}$O = 3.65).  
However, the more optically thin \CsvO\ emission ($\tau \sim 0.4$),
shown in the lower right of Figure 1b, displays similar behavior and 
does not increase for A$_V >$ 10 mag and 
we therefore interpret the structure seen in Figure 1a and 2a as the result 
of CO depletion.  A similar conclusion, using \thCO\ and \CeiO\ data,
has been reached by \citet{hotzel_b68}.

In Figure 2b we present the \NtwoHp\ J=1-0 and J=3-2 (2 detections) 
integrated intensity as a function of
\av .   
%The integrated emission from two detections of \NtwoHp\ J=3-2 
%are also shown as a function of \av\ to the lower right.   
In Figure 1b the
\NtwoHp\ emission exhibits a local minima at the location of the \av\ peak.
%and, thus
%is reminiscent of a smaller version of the CO depletion hole.  
In Figure 2b this minima is not clearly evident, and the positions in the depression
are located as a cluster of points with $\int T_{mb}$dv $\sim 2.25$ 
K km s$^{-1}$ for \av\ $>$ 25 mag.
A definitive conclusion regarding \NtwoHp\ depletion is complicated by
the presence of a few points at similar cloud depths with high intensities (points
located as the maxima on the ``ring'' in Figure 1b).  We find no difference
between the opacity derived from fits to the hyperfine structure from emission
in the ``ring'' and emission coincident with the depression.  
%Thus, the depression is not the result of increased optical depth.
However, due to the known presence of excitation gradients in B68 (ALL01) and the
expectation of significant non-LTE excitation for \NtwoHp ,
simple conclusions regarding depletion cannot be drawn from Figure 2b
\citep[e.g.,][]{bergin_ic5146}.

One additional notable feature in Figure 2b is the lack of significant \NtwoHp\ emission
for \av\ $<$ 5 mag.  A similar threshold is found in IC 5146
by \citet{bergin_ic5146}, which can be interpreted as a decrease in the
\NtwoHp\ abundance due to photodestruction of \Ntwo\ by the interstellar
radiation field (ISRF).

\subsection{Chemistry and Radiation Transfer}

To determine molecular abundances as a function of cloud depth
we have coupled the results from chemical models
to a 1-D spherical Monte-Carlo radiation transfer model
\citep{ashby_profile}.  
Our procedure is an iterative one in which we first estimate the 
radial abundance profile for a given
species from a time-dependent chemical model in which we fix profiles of the 
radial density, n$_{H_2}$($r$), and kinetic temperature, T$(r)$. 
The calculated abundance profile, along with the adopted density and 
temperature profiles, are then incorporated as input to the radiation 
transfer model (see Tf02).  The radiative transfer model determines the
expected profile of integrated intensity, which is convolved to the
appropriate angular resolution to compare directly to the observed data
\footnote{Similar to Tf02 for \NtwoHp , we calculate the level
excitation without accounting for hyperfine splitting, but include
the effects of the splitting
in the calculation of the emergent spectrum (for more discussion
see Tf02).}.
The parameters of the chemical model are 
then adjusted until the predicted integrated intensity profile 
matches the observed one.   For example, for CO the variables in 
this iterative solution
are: the magnitude of the incident local ISRF, 
the binding energy of CO to grains (effectively the desorption rate), and time.  

We fix the radial density profile used in the chemical and radiative transfer
models to that 
determined from the extinction profile (ALL01). 
Further, we adopt as the temperature profile that derived for dust in 
pre-stellar Bonnor-Ebert spheres by \citet{zucconi_tdust}, assuming 
T$_{dust} =$ T$_{gas}$. However,  we adjust the temperature
profile by a uniform reduction of 2~K in order to match the observed intensities
of both \NtwoHp\ transitions at the center of the cloud. 
Although the molecular data in B68 show evidence for systematic velocity 
gradients \citep{lada_b68}, we assume a static cloud in the radiative transfer model, which may 
overestimate the line opacities. However, we do incorporate the observed variation in 
the spectral line width in the model.  
The model line width includes contributions from the thermal and turbulent widths,
with the latter increasing as a function of radius.   
To be successful a model must reproduce both the dependence of $\int T$dv 
with \av\ (i.e., Figure 2) and $\Delta$v with radius (not shown).

%To place the data on the same plane as the observed quantities we
%derive the column density from the B68 radial density profile convolved 
%to a resolution of 25$''$, sampled 
%every 12$''$.  The model emission is then placed onto plots shown 
%in Figures 2a,b.  To compare the predicted velocity width
%to the data we fit all observations with a Gaussian.\footnote{For 
%\NtwoHp\ we used
%a multicomponent fit, constrained by the known hyperfine structure.}
%The observed FWHM were then radially averaged in 10$''$ bins.
 
%We are essentially fitting the observed spectra
%in a method similar to that performed by Tf02.

\subsubsection{C$^{18}$O}

For the chemistry of CO we adopted the time dependent analytical expression given
by \citet{caselli_ions}.  This model assumes that there 
is no gas phase CO formation and destruction, only depletion onto grain surfaces
and evaporation by cosmic ray impacts (using the formalism of \citet{hasegawa_cray}).
In the model, CO starts 
with an equilibrium abundance of $x$(CO) $= 8.5 \times 10^{-5}$ (relative
to H$_2$) and is allowed to adsorb and desorb.
Because we have information on the abundance at cloud edges we also account
for  \CeiO\ photodestruction using approximations of  
\citet{vDB88}.  We further adopted the standard ISRF \citep{habing68}. 
%\begin{equation}
%\[
%n(CO) = \frac{n(CO,\infty)}{C} + n(CO,\infty)[1 - \frac{1}{C}]exp(-\frac{t}{t_0})\;\;(1).
%\]
%\end{equation}
%\noindent Where $n(CO,\infty)$ is the steady state CO abundance of $x$(C0) $=$ 8.5 $\times 
%10^{-5}$, while
%$C$ and $t_0$ are combinations of the various 
%timescales involved in gaseous CO formation and destruction.
%These include depletion ($t_{dep}$) and cosmic ray desorption ($t_{cr}$) and
%the reader is referenced to \citet{caselli_ions} for more detailed
%descriptions ($C = 1 + \frac{t_{cr}}{t_{pd}}  + \frac{t_{cr}}{t_{dep}}$; 
%$t_0 = [t_{dep} t_{pd} + t_{cr} t_{dep} + t_{cr} t_{pd}]/t_{dep}t_{pd}t_{cr}$).

Due to the inclusion of \CeiO\ self-shielding, the
model is solved by lagging the solution of the CO column behind the
abundance determination. 
%Using the analytical expression we varied the time until the 
%model reproduced the \CsvO\ column
%density at the map center (N(C$^{17}$0 $= 2.2 \times 10^{14}$ cm$^{-2}$; 
%determined assuming LTE at 10 K).  
Similar to \citet{caselli_ions} and 
\citet{aikawa_dep} no solution is found consistent with a CO
binding energy, $E_b =$  960 K. Adopting 
$E_b =$  1210 K we find a solution at
t $\sim 6-7 \times 10^4$ years\footnote{As discussed in BL97 (and references therein)
a CO binding energy of 960 K is for CO bound to a CO ice covered surface, 
and 1210 K is representative
of CO frozen on a bare silicate surface.}.
The abundance profile derived from this model
is given in Figure 3 and the comparison to the observations is given in
Figure 2a.  To match the observed structure at low extinction required
reducing the UV field by adding an additional 1 mag of extinction to the model.  
This model provides
an excellent fit to the \CeiO\ data from \av\ $=$ 1 to 27 mag and suggests that
the \CeiO\ abundance declines by over 2 orders of magnitude from \av\ $= 1.5$ 
to the center.

\subsubsection{N$_{2}$H$^+$}

Because of the large number of potential gas phase reaction
partners for \NtwoHp\ we used the chemical model described in 
BL97.  
%This is a gas phase chemical model adapted to include depletion
%from the gas phase and cosmic ray desorption.
%In this model the density and extinction profile is fixed by the near-IR
%data and the chemistry allowed to evolve.  
The initial chemical abundances are set to evolved
cloud abundances taken from \citet{aikawa_disk}, except
$x$(CO) $=$ 8.5 $\times 10^{-5}$ and $x$(\Ntwo ) $= 2 \times 10^{-5}$.  
In addition we use $E_b$(CO) $= 1210$ K and $E_b$(\Ntwo ) $= 750$ K.  
This model is run until
the time matched the timescale determined by the CO analysis at which point
the theoretical CO abundance profiles predicted by the two
models were compared and found to be in good agreement.
  
%except at low extinction
%because of the difference in self-shielding between the isotopologues.
In this fashion the abundance of CO, a major destroyer of \NtwoHp , is fixed
and the primary variables that alter the \NtwoHp\ concentration are the abundance of
heavy metals and the binding energy 
of its pre-cursor \Ntwo .  Changing the metal ion
abundances will uniformly change the destruction rate of \NtwoHp\ at
depths $> 3$ mag (where carbon is not ionized).  In contrast, raising \Ntwo\ binding
energy will decrease the formation rate in high density regions
where depletion dominates. 

No match to the data given in Figure 2b is found for a constant \NtwoHp\ abundance.
An abundance reduction at cloud edges is required and 
the predicted emission does not fit the cluster of data points 
corresponding to the emission depression discussed in \S 3.1.  
% In addition no match is found for ``solar'' metal
%ion abundances.  
If we fix the metal ion abundances to the values
given by \citet{williams_ions}, we can match both the J $= 1-0$ and J $= 3-2$
emission, provided 
we lower the \citet{zucconi_tdust} 
temperature profile uniformly by 2~K (as mentioned earlier) 
and  raise the \Ntwo\ binding energy to 
E$_b$(\Ntwo ) $= 900$ K.  

The ``best'' match abundance profiles for B68 are given in Figure 3 and the 
fit to the observed emission is shown as a solid curve in Figure 2b.  The \NtwoHp\
abundance profile includes a reduction at cloud edges, 
peaks at \av\ $= 2$ mag with $x$(\NtwoHp ) $= 6 \times 10^{-11}$, 
and declines  by a factor of two by \av\ $ = 17$ mag. 
Over this range the concentration of \Ntwo\ declines by a factor of 5.
{\em The abundance profile, and N$_2$H$^+$ depletion, is constrained by
the observations and the radiative transfer;} however, the parameters of the 
chemical model may not be unique.   For example,
the \NtwoHp\ abundance is sensitive to
E$_b$(\Ntwo ) and the ionization fraction.
Observations of other molecular ions, which we are 
in the process of obtaining, will better constrain the ionization fraction
and the binding energy.  

%This time is essentially the ``depletion'' time
%required for CO to freeze out in an evolved cloud 
%(e.g. one that evolved at lower density without depletion
%and instantaneously reaches the observed density profile without chemical
%changes).  

\section{Discussion}

Our observations and analysis provide compelling evidence for
the existence of CO depletion in B68.
Perhaps the most interesting result of our experiment, however,
is the finding of \NtwoHp\ depletion.
The robustness of this result depends on the validity of our assumptions.
%and the precise \NtwoHp\ abundances may not be uniquely
%prescribed by our radiative transfer model. 
The density profile is well constrained by observations, but the temperature
structure is not.  B68 is a pre-stellar core; hence the most 
likely dependence is the
adopted one with hot edges and a cold center. 
Moreover, the detection of \NtwoHp\ J $= 3-2$ in the core center constrains
the temperature in the region where depletion is dominant.
Opacity is a concern.  The Monte-Carlo radiation transfer
properly handles the photon propagation, but we have not
accounted for the hyperfine structure or systematic line center velocity motions.
However, these effects would reduce the opacity, and re-enforce
the claim of depletion. Consequently the evidence for \NtwoHp\ depletion
in B68 is relatively firm.
In our best fit model we have chosen to fit between the scatter
of points at \av\  $> 20$ mag and not the points in the emission
depression.  Thus we likely have underestimated the \NtwoHp\ and
\Ntwo\ depletion.  Indeed it is entirely possible that \Ntwo\ is almost completely frozen
out in the dense center of B68, similar to CO.  
%Alternately if B68 is
%currently condensing, the depletion of \Ntwo\ will only increase. 

Depletion has been found to characterize the inner regions of many quiescent
cloud cores.   The chemical daughter
products of \Ntwo\ (\NtwoHp\ and \NHthree ) have previously been identified as the
best available tracers of these regions. 
However, our finding of \NtwoHp\ depletion
opens the possibility that the very centers of dense cores
could be effectively opaque to conventional methods of observation.
One casualty of this unfortunate situation is the search for infall wings
using optically thin molecular emission \citep{rawlings_dep} 
will be hampered by depletion in the very regions where the infall speed is highest. 
However, because \Htwo\ 
does not deplete onto grains, its chemical daughter
product, H$_2$D$^{+}$, could therefore prove to be the best tracer of gas in the innermost
regions of pre-stellar molecular cores.    
Indeed, because of \Ntwo\ and CO depletion, 
the abundance of H$_2$D$^{+}$ is expected to be large \citep{brown_deut}.
%As instruments are developed with greater sensitivity and
%new atmospheric windows are opened, H$_2$D$^{+}$ may become the best
%tracer in the use of molecular emission to probe the dynamics, chemistry,
%and physics of star formation.

Of course, B68 could be a unique object and may be more stable than cores like L1544.
This would allow for chemistry to evolve over longer times, resulting in 
greater freeze out.  Alternatively, \citet{charnley_n2} showed that selective
depletion could lead to \Ntwo\ destruction by gas-phase chemistry, primarily 
at longer timescales and higher densities.
Because of the potential implications of near complete
molecular freeze-out, additional searches for
\NtwoHp\ and \NHthree\
depletion are needed to infer the general applicability of this result.  
Regardless, our observations and analysis provide important constraints on the 
chemical evolutionary processes active in the B68 cloud.

\acknowledgements

We are grateful to the referee, Steve Charnley, for helpful comments.
This work was supported by Grant \#NAG5-9520 from NASA's Origins Program.

%\begin{deluxetable}{lllccc}
%\tablenum{1}
%\tablewidth{5in}
%\tablecolumns{7}
%\tablecaption{Observed Transitions and Telescope Parameters}
%\tablehead{
%\colhead{Molecule} &
%\colhead{Transition } &
%\colhead{$\nu$(GHz)} &
%\colhead{$\theta_{MB}$($''$)} &
%\colhead{$\Delta v_{res}$ (km s$^{-1}$)} &
%\colhead{$\sigma$ (K)} \\ 
%}
%\startdata
%C$^{18}$O & J $=$ 1 -- 0 & 109.78218 & 22 & 0.053 & 0.14 \\ 
%C$^{17}$O & J $=$ 1 -- 0 & 112.35928 & 21 & 0.104 & 0.05 \\ 
%N$_2$H$^+$ &J $=$ 1 -- 0 & 93.17378 & 26 & 0.126 & 0.08 \\
%N$_2$H$^+$ & J $=$ 3 -- 2 & 279.51194  & 10 & 0.021\tablenotemark{a} & 0.08 \\
%\enddata
%\tablenotetext{a}{To improve the signal to noise these data were smoothed to 
%$\Delta v_{res} = 0.168$  km s$^{-1}$.}
%\end{deluxetable}
%

\begin{center}
\begin{figure*}
\includegraphics[height=8cm]{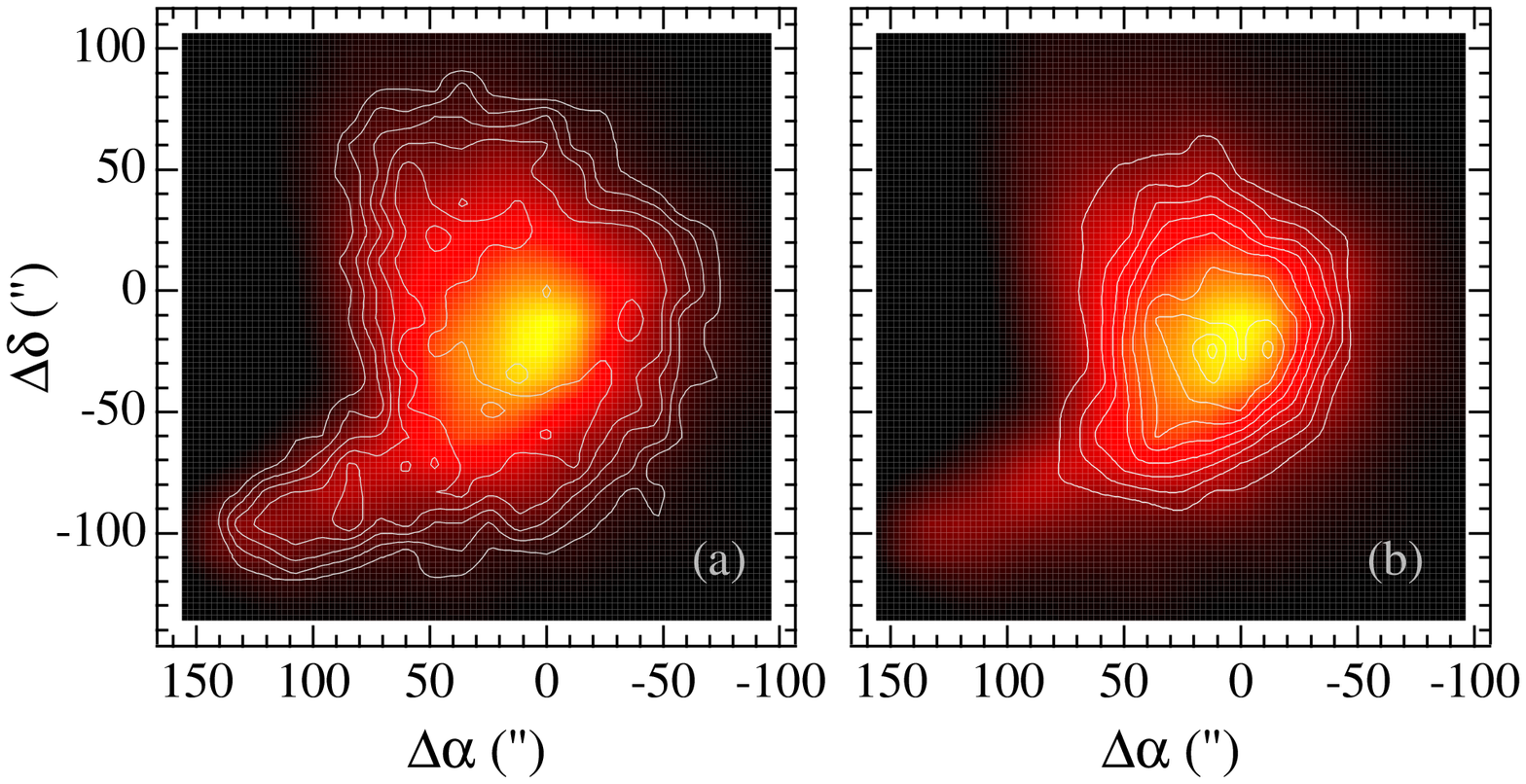}
\caption{
(a) Comparison of \CeiO\ J=1-0 integrated emission (contours) in B68 superimposed on a
map of visual extinction derived by ALL01 and convolved to the IRAM resolution.  
The \CeiO\ contours
begin at 0.2 K km s$^{-1}$ and step in units of 0.1 K km s$^{-1}$.  The
A$_V$ image scales is scaled from  0--27 mag. (b) Comparison of 
\NtwoHp\ J=1-0 integrated emission (contours) and the  
visual extinction image.  The \NtwoHp\ contours begin at 0.3 K km s$^{-1}$
and step in units of 0.2 K km s$^{-1}$.  
These maps were referenced to $\alpha$ = 17$^h$22$^m$38\fs 2
and $\delta = -$23$\arcdeg$49$'$34\farcs 0 (J2000).
The peak A$_V$ is located 6$''$ east and 12$''$ south of this position.
}
\end{figure*}
\end{center}
\begin{center}

\begin{figure*}
%\includegraphics[height=7cm]{f2.eps}
%\includegraphics[height=21cm]{b68_fig2.eps}
%\includegraphics[height=6.5cm]{x.eps}
%\plotone{b68_fig2.eps}
\plotone{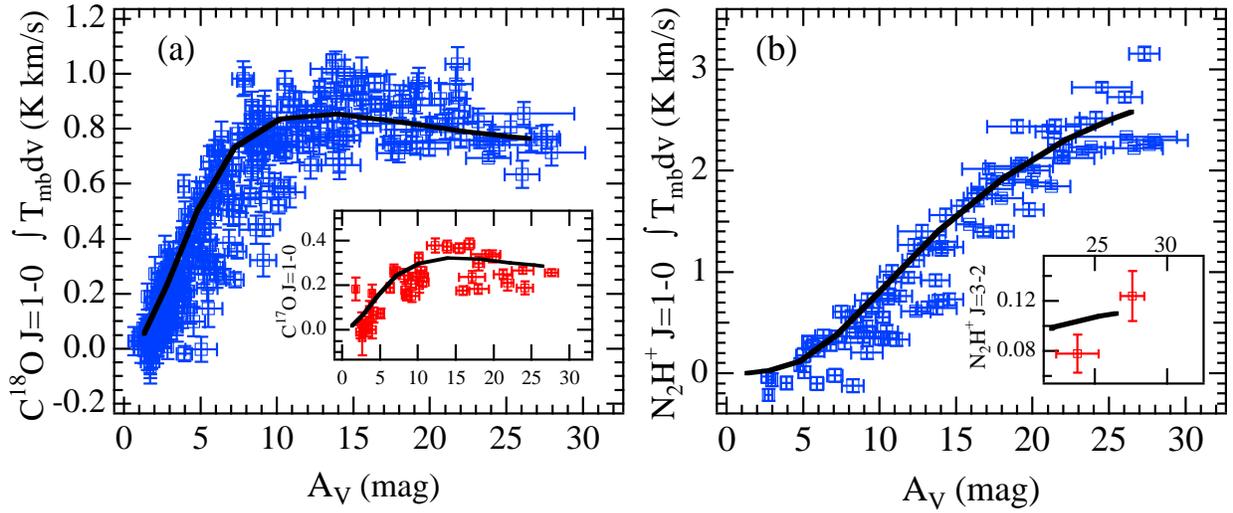}
\caption{
(a) C$^{18}$O J=1-0 integrated intensity as a function of visual extinction for
the entire B68 dark cloud.  Inset to the lower right is a similar plot for
\CsvO . (b)  N$_2$H$^+$ J=1-0 integrated emission as a function
of visual extinction in B68.  Inset to the lower right are the integrated intensities
(in K km s$^{-1}$) of the 2 \NtwoHp\ J=3-2 detections at the corresponding extinctions.
In all plots the data are presented as open
squares with error bars while solid curves represent the emission predicted by
a model combining chemistry with a Monte-Carlo radiative transfer code
(\S 3.2). To create these figures the visual 
extinction data was convolved to a angular resolution of 25$''$ and
sampled on the same grid as the molecular observations.  For  \NtwoHp\ J=3-2 the
molecular data and model both have an angular resolution of 10$''$.
}
\end{figure*}
\end{center}

\begin{figure*}
\includegraphics[height=25cm]{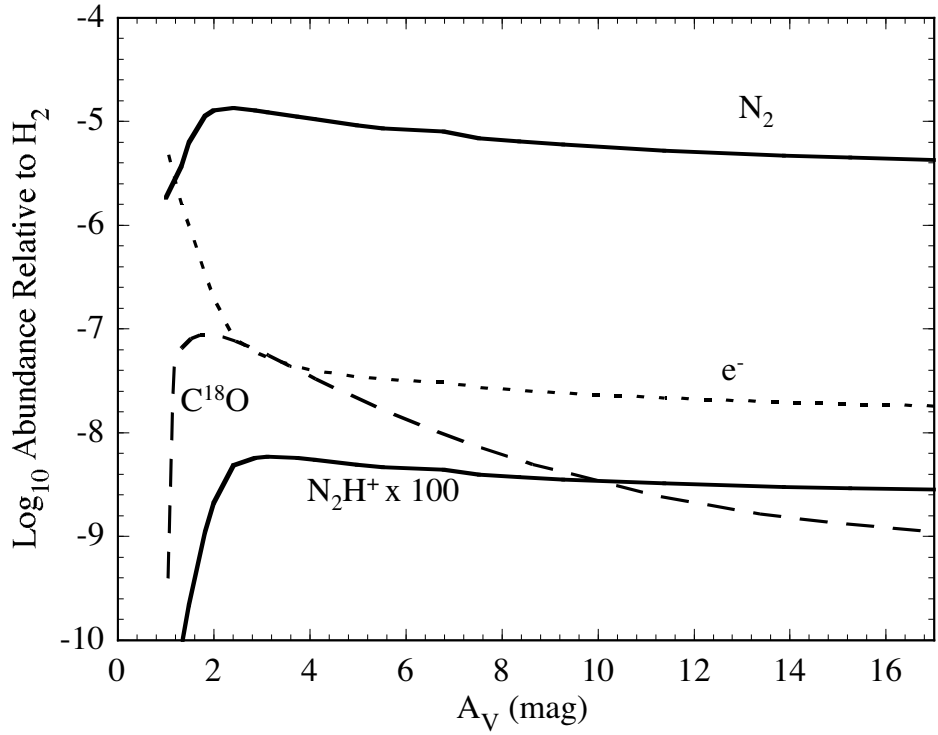}
\vspace{-5.0in}
%\epsscale{0.80}
%\plotone{b68_fig3_2.eps}
\caption{
Molecular abundances relative to H$_2$ as a 
function of visual extinction for the best fit model of B68.
The abundance of \NtwoHp\ declines by a factor of 2 from A$_V = 2$ mag to 
A$_V = 17$ mag.
}
\end{figure*}
\end{document}